# First-principles theory of extending the spin qubit coherence time in hexagonal boron nitride

Jaewook Lee, Huijin Park, and Hosung Seo*

Department of Physics and Department of Energy Systems Research, Ajou University, Suwon, Gyeonggi 16499, Korea

**ABSTRACT**

Negatively charged boron vacancies ($V_B^-$) in hexagonal boron nitride (h-BN) are a rapidly developing qubit platform in two-dimensional materials for solid-state quantum applications. However, their spin coherence time ($T_2$) is very short, limited to a few microseconds owing to the inherently dense nuclear spin bath of the h-BN host. As the coherence time is one of the most fundamental properties of spin qubits, the short $T_2$ time of $V_B^-$ could significantly limit its potential as a promising spin qubit candidate. In this study, we theoretically proposed two materials engineering methods, which can substantially extend the $T_2$ time of the $V_B^-$ spin by four times more than its intrinsic $T_2$. We performed quantum many-body computations by combining density functional theory and cluster correlation expansion and showed that replacing all the boron atoms in h-BN with the $^{10}B$ isotope leads to the coherence enhancement of the $V_B^-$ spin by a factor of three. In addition, the $T_2$ time of the $V_B^-$ can be enhanced by a factor of 1.3 by inducing a curvature around $V_B^-$. Herein, we elucidate that the curvature-induced inhomogeneous strain creates spatially varying quadrupole nuclear interactions, which effectively suppress the nuclear spin flip-flop dynamics in the bath. Importantly, we find that the combination of isotopic enrichment and strain engineering can maximize the $V_B^-$ $T_2$, yielding 207.2 and 161.9 μs for single- and multi-layer h-$^{10}$BN, respectively. Furthermore, our results can be applied to any spin qubit in h-BN, strengthening their potential as material platforms to realize high-precision quantum sensors, quantum spin registers, and atomically thin quantum magnets.

**INTRODUCTION**

Optically addressable spin defects in wide band-gap materials are promising solid-state qubit platforms that enable cutting-edge quantum applications such as quantum computation [1], quantum sensing [2], and quantum network [3,4]. Spin defects in diamond [5], which show long spin coherence times and many other attractive features such as high temperature quantum functionality [6] and spin-to-photon interfaces [7], is one of the leading qubit systems for quantum applications. Progresses made in the research on diamond have inspired several pioneering studies, wherein quantum spin defects have been developed in silicon carbide and nitrides, broadening the palette of spin qubits in materials [8-11]. Notably, the spin qubits in non-diamond hosts provide unique opportunities for building advanced platforms for quantum systems by taking advantage of the well-established material technologies developed for their hosts [12-13]. Recently, the search for qubit systems in two-dimensional (2D) van der Waals (vdW) materials has gained significant attention owing to their potential superiority in light extraction, heterostructuring, defect positioning, strain engineering, and nano-photonic integration [14-17].

Among the 2D vdW materials, quantum spin defects in hexagonal boron nitride (h-BN) are gaining prominence for the development of optically active spin qubits. Owing to its wide bandgap of ~6 eV, h-BN hosts a variety of color centers from near infrared to ultraviolet [15,16,18]. Some of the color centers have been found to emit bright single photons even at room temperature, and these findings have sparked worldwide research efforts in this direction [19-25]. Subsequently, the search for optically addressable spin qubits in h-BN has become the focus of materials research, and several breakthroughs have been made recently. Notably, negatively charged boron vacancies ($V_B^-$) were discovered as optically addressable spin qubits in h-BN [26-28]. Since the first report by Gottscholl *et al.*, [26] several important achievements have been made [29], including the realization of coherent Rabi oscillations [30,31], deterministic defect generations [32-34], nano-scale quantum sensing [35], and coupling to nano-photonic structures [36]. In addition, carbon-related defects [37-39] and boron vacancy complexes [40] have also been recently identified as optically accessible spin qubits in h-BN. Overall, significant advances have been made in the development of defect-based spin qubits in h-BN and their use in quantum applications. Nonetheless, further research is required to realize robust spin qubits in h-BN.

One of the most compelling issues for h-BN spin qubits is their short spin coherence time ($T_2$) due to the dense nuclear spin bath in the h-BN lattice [26,40,41]. For $V_B^-$, the Hahn-echo $T_2$ time has been measured to be several microseconds [30,31]. Notably, several well-known schemes, such as isotopic purification [42], dynamical decoupling (DD) [43], and clock transitions (CT) [44,45], are available to extend the spin coherence times in materials. However, in h-BN, a conventional isotopic purification is impossible because all the naturally occurring boron ($^{10}$B and $^{11}$B) and nitrogen ($^{14}$N and $^{15}$N) isotopes exhibit non-zero nuclear spins. In addition, the coherence protection schemes such as DD and CT are often limited by the intrinsic Hahn-echo $T_2$ time. When the $T_2$ time is extremely short like in h-BN, pulse requirements in the DD and CT schemes become considerably challenging [11]. Considering the fundamental role of the coherence time in determining the retention time of the quantum information, the short $T_2$ time of the h-BN spin qubits is one of the most pressing problems that needs to be solved for advancing these platforms in h-BN.

In this study, by taking the $V_B^-$ spin as a representative spin qubit system in h-BN, we developed two unconventional methods to enhance the $T_2$ time of spin qubits significantly in h-BN via isotopic enrichment and strain engineering. We combined density functional theory (DFT) and cluster correlation expansion (CCE) to compute the theoretical $T_2$ time of $V_B^-$ in h-BN single layer and multiple layers to be 45.9 and 26.6 μs, respectively, in the presence of an intrinsic nuclear spin bath in natural h-BN. Subsequently, we showed that the $T_2$ times were increased to 143.4 and 81.1 μs in the single- and multi-layer h-$^{10}$BN, respectively, which is enriched with the $^{10}$B isotope. This result is somewhat counterintuitive as the $^{10}$B isotope has a larger nuclear spin ($I = 3$) than that of the $^{11}$B isotope ($I = 3/2$). We found that this $T_2$ enhancement by a factor of three resulted from the smaller gyromagnetic ratio of $^{10}$B than that of $^{11}$B, despite the much larger nuclear spin of $^{10}$B. In addition, we demonstrated that the $T_2$ time could be further increased to 207.2 and 161.9 μs for the single- and multi-layer h-$^{10}$BN, respectively, by inducing inhomogeneous strain around the $V_B^-$. We considered a Gaussian-shaped bubble around $V_B^-$ as a representative inhomogeneous strain and showed that the $T_2$ increased in proportion to the Gaussian width and height, demonstrating the impact of the local inhomogeneous strain on the $T_2$ time. We found that a spatial inhomogeneity in the nuclear spin quadrupole interaction effect, induced by the inhomogeneous strain, plays a crucial role in suppressing the nuclear spin flip-flop dynamics, thus enhancing the $V_B^-$ spin coherence. Our results pave the way for effective protection of the spin coherence in h-BN, and this method should be applicable to $V_B^-$ as well as to any potential spin qubits in h-BN [40,46,47]. Improved spin coherence of $V_B^-$ would also open possibilities of further coherence extension using DD or CT, thereby advancing spin qubits in h-BN as promising platforms for quantum information science and technology.

**RESULTS AND DISCUSSION**

**Spin decoherence of $V_B^-$ in natural h-BN.** We considered a central spin model to compute the decoherence dynamics of a $V_B^-$ spin interacting with the nuclear spin bath in the h-BN lattice. Fig. 1. (a) shows a schematic of the spin model, in which a $V_B^-$ defect is created in the middle of a large h-BN supercell. The ground state of $V_B^-$ is a spin-triplet state, and the spin density is highly localized at the vacancy site (see inset of Fig. 1. (a)). In our model, we assumed a localized $S = 1$ at the vacancy site and treated its +1 and 0 spin sublevels as qubit states. The nuclear spin bath, which strongly interacts with the $V_B^-$ spin via hyperfine interaction, is derived from the spin-bearing boron and nitrogen nuclei in h-BN. Notably, all the naturally occurring boron and nitrogen isotopes exhibit non-zero nuclear spins: 19.9 % of $^{10}$B with $I = 3$, 80.1 % of $^{11}$B with $I = 3/2$, 96.6 % of $^{14}$N with $I = 1$, and 0.4 % of $^{15}$N with $I = ½$. In our model, we randomly distributed the nuclear spins in the h-BN lattice according to their natural abundance. To compute the homogeneous dephasing time ($T_2$) of the $V_B^-$ spin, we considered the Hahn-echo pulse sequence [48] and used the CCE method to expand the many-body nuclear-spin correlation effects systematically on the $V_B^-$ spin decoherence [49]. Notably, the CCE method combined with DFT calculations enables the quantum many-body computation of $T_2$ without the assumption of any adjustable theoretical parameter. Owing to its predictive power, the CCE method was successfully applied to a wide range of solid-state qubit systems, yielding excellent agreement with the experimental results [11,50-54] (further details on the theoretical methods and the system's spin Hamiltonian are provided in the Method and Supplementary Information sections; see Supplementary Note 1 and 2, and Figure S1).

Fig. 1. (b) presents the computed spin coherence of $V_B^-$ in the bulk of natural h-BN. Evidently, the spin coherence rapidly decays during the free evolution time, within tens of microseconds. By fitting the coherence with a stretched exponential function ($\exp(-t_{free}/T_2)^n$), we computed the $T_2$ time and stretching exponent (*n*) of the decay to be 26.64 µs and 2.64, respectively. Additionally, we determined $T_2$ and *n* of the $V_B^-$ spin in the single-layer h-BN to be 45.85 µs and 2.27, respectively, owing to the reduced number of nuclear spins surrounding $V_B^-$ (see Figure S2.). Notably, the computed $T_2$ time corresponds to the upper limit of $T_2$ set by the intrinsic nuclear spin bath of h-BN. The experimentally measured $T_2$ time can be smaller than the theoretical $T_2$ [30,40], if other decoherence sources such as other paramagnetic defects are present in h-BN.

To examine the nuclear bath dynamics generating the intracrystalline magnetic noise, we analyzed the impact of the spin Hamiltonian terms on the $V_B^-$ spin decoherence. Importantly, we find that the nuclear quadrupole interaction [55] plays an important role in determining the bath dynamics in h-BN. Fig. 1. (b) shows the $V_B^-$ decoherence computed with a partial spin Hamiltonian, in which the quadrupole Hamiltonian terms are excluded from the model. We observe that the spin coherence decays much faster in the "without-the-quadrupole" model than in the "with-the-quadrupole" model. Without the quadrupole interaction effect in the model, the $T_2$ time of $V_B^-$ is reduced to 17.92 and 34.98 μs for the multi- and single-layer h-BN, respectively [41]. Notably, the nuclear spin flip-flop transitions, driven by the magnetic dipolar coupling between the nuclear spins, are the dominant sources of intracrystalline magnetic noise in a nuclear spin bath [50,54,56]. Our results show that the nuclear quadrupole interaction in h-BN plays a significant role in suppressing such nuclear flip-flop transitions in the h-BN.

To understand the role of quadrupole interaction in the pairwise nuclear spin transitions, we show the energy levels of two $^{10}$B nuclear spins, defined as $|m_{I_1}, m_{I_2}\rangle$, where $m_I$ = 3, 2, 1, 0, -1, -2, -3, in Fig. 1. (c)–(e). Fig. 1. (c) shows the energy level splitting due to the Zeeman interaction ($H_Z$) and hyperfine field imposed by the $V_B^-$ spin ($H_{HF}$). The Zeeman splitting yields spin manifolds, which are separated from each other in energy by 13.7 MHz in the presence of an external magnetic field of 3 T. Each manifold contains certain number of two-spin states such as $\{|-3,-1\rangle, |-2,-2\rangle, |-1,-3\rangle\}$, which have the same Zeeman energy, as shown in Fig. 1. (d). These states in a manifold are split by a small energy of the order of hundreds of Hz because of the hyperfine field (denoted as $\Delta_{HF}$ in Fig. 1. (d)). In the "without-the-quadrupole" decoherence model, the magnetic dipolar interaction drives the flip-flop transitions between the states, confined in each manifold, with a transition rate of hundreds of Hz (denoted as $\Omega_{FF}$ in Fig. 1. (d)). However, the transitions are strongly suppressed between the states belonging to different manifolds because of a large Zeeman energy mismatch. Remarkably, the energy levels in a manifold exhibit significant splitting in the presence of the quadrupole interaction, as shown in Fig. 1. (e). We found that the quadrupole-induced splitting ($\Delta_{HF+Q}$ in Fig. 1. (e)) ranges from tens of kHz to a few MHz, owing to the large nuclear quadrupole interaction in h-BN (see Table 1). Thus, the flip-flop transitions within a Zeeman manifold are significantly suppressed, compared to those evaluated using the "without-the-quadrupole" model, because of the large quadrupole-driven energy mismatch. Our analysis reveals the importance of the quadrupole interaction in determining the nuclear bath dynamics, which in turn governs the spin decoherence of the spin qubits in h-BN.

**Isotopic enrichment of h-BN to increase $T_2$.** Fig. 2 depicts the computed $T_2$ of $V_B^-$ in h-BN as a function of the composition ratio of $^{10}$B and $^{14}$N in h-BN. Surprisingly, we find that the $T_2$ time substantially increases as the ratio of $^{10}$B and $^{14}$N increases toward 100 % in the lattice, despite their larger nuclear spins than those of $^{11}$B and $^{15}$N. When h-BN is 100 % enriched with $^{10}$B and $^{14}$N (i.e., h-$^{10}$B$^{14}$N), the $T_2$ time is increased to 143.39 and 81.11 μs in the single- and multi-layer h-$^{10}$B$^{14}$N, respectively. These values are three times larger than those in natural h-BN. In contrast, the $T_2$ times in single- and multi-layer h-$^{11}$BN are 45.85 and 26.64 μs, respectively, which are unanticipated, because a smaller nuclear spin (e.g., $I = 3/2$ of $^{11}$B vs. $I = 3$ of $^{10}$B) results in a smaller number of flip-flop transition channels in the nuclear spin–spin interaction as well as lower intracrystalline noise in h-$^{11}$B$^{15}$N than in h-$^{10}$B$^{14}$N. However, our results show that the opposite is true—the intracrystalline magnetic noise is significantly reduced in h-$^{10}$B$^{14}$N.

To elucidate the microscopic origin of the enhanced $T_2$ in h-$^{10}$B$^{14}$N, we consider a hypothetical central spin model, wherein the gyromagnetic ratio ($\gamma_B$) and the nuclear spin number ($I_B$) at the boron sites are variable. Fig. 3 shows $T_2$ as a function of $\gamma_B$ and $I_B$. Considering the actual $I_B$ and $\gamma_B$ of $^{10}$B and $^{11}$B (2.875 $\frac{rad}{G \cdot ms}$ and 8.585 $\frac{rad}{G \cdot ms}$ [58], respectively), we consider the range of $\gamma_B$ to be from 1 to 10 $\frac{rad}{G \cdot ms}$, and that of $I_B$ from ½ to 3 in Fig. 3. We find that for a given $\gamma_B$, the $T_2$ time decreases as $I_B$ increases, indicating that the intracrystalline magnetic noise increases as the total number of nuclear spin flip-flop channels in the bath (proportional to $(2I_B)^2$) increases. Additionally, $T_2$ increases quadratically as $\gamma_B$ decreases. For instance, for $I_B$ = 3/2, the $T_2$ changes from 24 to 695 μs when $\gamma_B$ changes from 10 to 1 $\frac{rad}{G \cdot ms}$. The $\gamma_B$-dependent change in $T_2$ observed mainly because the magnetic dipolar coupling strength is proportional to $(\gamma_B)^2$. Thus, decreasing $\gamma_B$ reduces the flip-flop transition rate quadratically in the nuclear spin bath. In h-$^{10}$BN, the total number of flip-flop transitions is increased by four times, but the flip-flop transition rate is substantially decreased by approximately nine times compared to those in h-$^{11}$BN. In the case of $^{14}$N isotope, the same analysis yields four times larger number of flip-flop transition channels and approximately two times smaller flip-flop transition rates in h-B$^{14}$N compared to those in h-B$^{15}$N. Therefore, based on the result shown in Fig. 3, we can conclude that in h-BN, the spin coherence of $V_B^-$ is much more sensitive to the change in $\gamma_B$ than that in $I_B$, and the increased $T_2$ in h-$^{10}$B$^{14}$N stems from the smaller $\gamma_B$ of the $^{10}$B and $^{14}$N nuclear spins despite their large nuclear spin numbers.

**Effects of inhomogeneous lattice strain on spin coherence.** Next, we investigate the impact of inhomogeneous lattice strain on the $T_2$ of $V_B^-$ spin in h-BN. We consider a curved h-BN to create an inhomogeneous strain around $V_B^-$. Reportedly, various feasible methods have been developed to study the curvature-induced phenomena in h-BN by creating bubbles [59,60], pillars [34], folds [61], and wrinkles [20]. As a representative model of the local curvature and inhomogeneous strain, we consider a Gaussian lattice deformation as illustrated in Fig. 4. (a). The lattice deformation can be described in terms of the full width at half maximum (FWHM) and height of the Gaussian function, and the $V_B^-$ defect is created on top of the Gaussian deformation. We employed DFT to compute the spin Hamiltonian parameters in the deformed h-BN lattice. The computed effects were subsequently included into the spin Hamiltonian to examine the effect of the inhomogeneous lattice strain on the spin coherence of $V_B^-$.

Fig. 4. (b) shows the computed $T_2$ time of the $V_B^-$ in natural h-BN (no isotopic modification) as a function of the FWHM and height of the Gaussian lattice deformation. Evidently, the $T_2$ time increases significantly with the increasing function height and FWHM. At a function height of 3.5 Å, the $T_2$ in the single- and multi-layer h-BN is 58.7 and 34.2 µs, respectively, as shown in Fig. 4. (b) and (c). These values are 12.9 and 7.5 µs greater than those in natural and flat h-BN.

To understand the curvature-induced enhancement of the $T_2$ time, we analyze the impact of the lattice deformation on the $V_B^-$ spin decoherence in terms of two factors: modification of the nuclear spin–spin distances and spatial inhomogeneity induced in the nuclear spin quadrupole interaction in the deformed h-BN lattice. In Figure S3, we consider a model that does not include the nuclear spin quadrupole interaction, and compute $T_2$ as a function of FWHM and function height same as before (Fig. 4). We find that the $T_2$ time shows negligible changes, as shown in Figure S3, implying that the enhanced $V_B^-$ spin coherence, obtained using the Gaussian lattice deformation, is not derived from the mere modification of the nuclear spin–spin distances; instead, it is mediated by the modified nuclear quadrupole interaction produced by the inhomogeneous strain.

To visualize the role of nuclear quadrupole interaction in determining the strain-driven enhancement of $T_2$, we compute the first-order quadrupole-interaction-induced energy shift ($\zeta$) of a nuclear spin in h-BN. As described in Supplementary Note 3, the energy shift is calculated as: $Q_C \times \zeta \times (3m_I^2 - I(I+1))$, where $Q_C$ is the coefficient of quadrupole interaction (6), and $\zeta = V_{zz} - (V_{xx}+V_{yy})/2$, where $V_{ii}$ (I = x, y, z) is the electric field gradient computed using DFT. Fig. 5. (a) to (c) show $\zeta$ as a function of position with respect to the $V_B^-$ site, indicating that a difference of $\zeta$ between two different points, shown in Fig. 5. (a)–(c), is proportional to the difference between the quadrupole-induced energy shifts of the two nuclear spins at these two different points. As shown in Fig. 5. (a) to (c), the inhomogeneity of $\zeta$ becomes broader and larger as the Gaussian lattice deformation increases. In terms of the $^{10}$B–$^{10}$B two-spin energy levels presented in Fig. 1. (e), a larger difference of $\zeta$ between two $^{10}$B spins results in a larger level splitting in their two-spin energy levels.

Nuclear spin flip-flop transitions are well understood in terms of a pseudospin model [50,54], which can model a flip-flop transition between any two $^{10}$B–$^{10}$B nuclear spin states having the same Zeeman energy. The two states of a pseudospin are separated by an energy gap ($\Delta$), and the transition between them is mediated by the magnetic dipolar interaction with a flip-flop transition rate ($\Omega$) (see Supplementary Note 4). Fig. 5. (d)–(e) show the histogram of $(\log(\Omega/\Delta))$ for all the possible pseudospins in the Gaussian-deformed h-BN. Notably, a small $\log(\Omega/\Delta)$ indicates a high suppression in the flip-flop transition. Furthermore, a large Gaussian deformation around the $V_B^-$ defect produces a large number of pseudospins with small $\log(\Omega/\Delta)$ values appearing in the histogram. Our results demonstrate that the energy gap between numerous $^{10}$B–$^{10}$B pseudospins in h-BN increases as the Gaussian lattice deformation becomes larger, which in turn produces a larger inhomogeneous strain around $V_B^-$. This results in a considerable suppression of the nuclear flip-flop transitions in h-BN, leading to the enhancement of the $V_B^-$ $T_2$ time.

**Maximizing the spin coherence time.** The two methods suggested in this study can be combined to maximize the $T_2$ time. Fig. 6. (a) and (b) present the computed $T_2$ as a function of the Gaussian height and FWHM for the $V_B^-$ in the Gaussian-deformed h-$^{10}B^{14}N$. Evidently, the maximum achievable $T_2$ is 207.17 and 161.92 μs in the single- (Fig. 6. (c)) and multi-layer h-BN (Fig. 6. (d)), respectively. These values are approximately ~ six times larger than that in natural and flat h-BN. Particularly, the combined effect of the smaller gyromagnetic ratio of $^{10}B$ and the inhomogeneous quadrupole interaction induced by the inhomogeneous strain leads to the enhancement of $T_2$ of the $V_B^-$ spin.

**CONCLUSION**

In summary, we proposed two theoretically effective methods to extend the spin coherence time of the $V_B^-$ spin qubit in h-BN. The spin coherence time is otherwise limited to a few microseconds because of the strong intracrystalline magnetic noise caused by the dense nuclear spin bath in the h-BN lattice. To predict the $T_2$ time of the spin qubit accurately, we performed first-principles calculations by combining the DFT and CCE theory. In a natural h-BN host, the theoretical upper limit of the $V_B^-$ $T_2$ time is 45.85 and 26.64 μs in single- and multi-layer h-BN, respectively. Next, we demonstrated that the $T_2$ time can be increased to 143.39 and 81.11 μs in single- and multi-layer h-$^{10}BN$, respectively, in which all the boron atoms in the lattice are replaced by the $^{10}B$ isotope. It is evident that such an isotopic enrichment technique has been already developed and applied to h-BN [62,63]. By analyzing the magnetic dipolar interaction between the boron nuclear spins, we showed that the smaller gyromagnetic ratio of $^{10}B$ plays a key role in suppressing the nuclear flip-flop dynamics, despite its large nuclear spin compared to that of $^{11}$B. Then, we showed that the $T_2$ time of $V_B^-$ can be further increased by introducing an inhomogeneous strain around the $V_B^-$ defect. Applying a Gaussian-type lattice deformation (as a representative model of the inhomogeneous strain), we identified that the inhomogeneous strain contributes to the suppression of the nuclear spin flip-flop dynamics in the bath by producing spatially varying nuclear quadrupole interactions in h-BN. We found that applying both the isotopic enrichment and inhomogeneous strain could increase the $V_B^-$ $T_2$ time by six times that in a pristine h-BN bulk, and consequently, the $T_2$ can reach up to 207.17 and 161.92 μs for the single- and multilayer h-BN, respectively.

Our study on $V_B^-$ not only provides a fundamental understanding of the decoherence of $V_B^-$ spins in h-BN but paves the way to engineer their $T_2$ times. With an increased $T_2$ time, the $V_B^-$ spin coherence could be further enhanced by combining other active quantum control schemes such as DD or CTs. In addition, the essential physics developed in this study can be applied to any localized spin in any 2D vdW material. Thus, the proposed approach provides a universal tool for engineering the coherence of potential spin qubits in 2D vdW materials. Combining these engineering schemes with the unique characteristics of 2D vdW materials would be a promising strategy for the development of robust low-dimensional quantum systems.

***Erratum added:*** We recently found that there was a mistake in the $T_2$ calculation of the $V_B^-$ spin in the Gaussian-deformed multi-layer h-BN lattice. It was found that the quadrupole interactions in h-BN layers adjacent to the one embedding the $V_B^-$ defect were not correctly treated. To be more specific, we used DFT to compute the quadrupole interactions in the presence of the Gaussian lattice deformation around $V_B^-$, which are transferred to subsequent CCE computation. The mistake was found in the data transfer process. We found that the quadrupole interaction computed for strain-free h-BN layer was used for some Gaussian-deformed h-BN layers near the $V_B^-$ defect. This mistake affected the results of Figure 4 (c) and Figure 6 (b) and (d), which are corrected in the current version of the manuscript. In the corrected result, the $T_2$ time in the Gaussian-deformed multi-layer h-$^{10}$BN reaches 161.92 μs.

## METHODS

**Quantum-bath approach for decoherence and CCE.** According to the quantum-bath theory, the spin qubit decoherence occurs because of the entanglement between the qubit and its environment [64]. In our spin model, the $V_B^-$ spin serves as a qubit, and the main environmental degrees of freedom are the nuclear spins residing in the h-BN lattice. The nuclear spins are coupled to each other via magnetic dipolar coupling, and the nuclear spin bath interacts with the qubit through hyperfine interactions. The full spin Hamiltonian for the entire quantum many-body system is described in Supplementary Note 1. To compute the homogeneous dephasing time—$T_2$—of the $V_B^-$ ensembles, the Hahn-echo pulse sequence, which features a $\pi$ pulse between two free evolution time $\tau$, is considered. The decoherence of the qubit is obtained by computing the off-diagonal elements of the reduced density matrix of the qubit after tracing out the bath degrees of freedom at the end of the $2\tau$ free evolution time, as: $\mathcal{L}(2\tau) = \frac{\text{Tr}[\rho(2\tau)S_+]}{\text{Tr}[\rho(0)S_+]}$, where $S_+$ is the electron spin raising operator, and $\rho$ is the density operator of the entire system. To compute the coherence function, we employed the CCE method [49,65], which systematically expands the coherence function. We find that the CCE expansion converges at the CCE-2 level of the theory, implying that the nuclear spin–spin pairwise correlation effect is the dominant source of decoherence. Extensive numerical convergence tests were performed in terms of the CCE order, bath size, and coupling radius, and the corresponding results are summarized in Supplementary Note 2 and Figure S1.

**DFT calculations.** We performed DFT calculations using plane-wave basis functions with an energy cutoff of 85 Ry, as implemented in the Quantum Espresso (QE) code [66], to compute the nuclear quadrupole interaction and the contact hyperfine interaction in h-BN containing $V_B^-$ with a Gaussian lattice deformation. We used the Perdew–Burke–Ernzerhof exchange-correlation functional along with the optimized norm-conserving Vanderbilt pseudopotentials [67,68] and projector-augmented wave pseudopotentials [69,70]. To simulate an isolated $V_B^-$ defect in the h-BN, we used large supercells containing either single- or multi-layer h-BN. For the single-layer h-BN calculations, we built a 336-atom supercell starting from a 4-atom orthorhombic unit cell. The h-BN layer in the supercell was separated from its periodic images in the out-of-plane direction by a 10-Å-thick vacuum space. For the multi-layer h-BN, a 672-atom supercell was used. To compute the electric field gradient tensor and the contact hyperfine interaction, we employed the GIPAW module as implemented in the QE code [71].

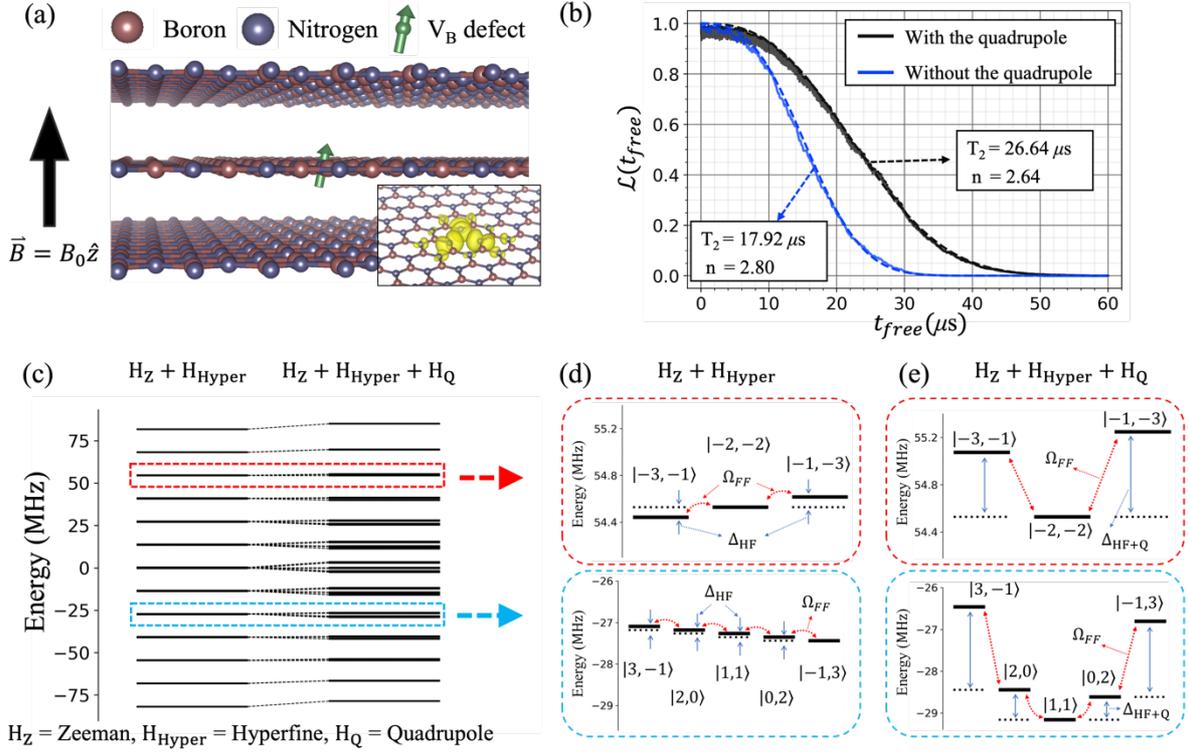

**Figure 1**. **Spin decoherence of $V_B^-$ in bulk h-BN.** (a) Schematic of the $V_B^-$ spin in multi-layer h-BN. An external magnetic field ($B_0$) of 3 T is applied along the out-of-plane direction. Inset shows the ground-state spin density of $V_B^-$, computed using DFT. (b) Spin coherence of the $V_B^-$ spin in multi-layer h-BN (black line). The spin coherence is fitted with a stretched exponential function, $\exp(-t_{free}/T_2)^n$, to compute the spin coherence time ($T_2$) and stretching exponent ($n$). A hypothetical model of the bath in which the nuclear quadrupole interaction is ignored was considered for comparison (blue line). (c) Energy levels of two $^{10}B$ nuclear spins in the presence of Zeeman effect ($H_Z$) and a hyperfine field ($H_{Hyper}$) imposed by the $V_B^-$ electron spin (left) and in the presence of $H_Z$, $H_{Hyper}$, and the nuclear quadrupole interaction ($H_Q$) (right) (d, e) Detailed energy-level diagrams for the two specific manifolds, denoted by the red and blue dotted lines in (c), (d) in the presence of $H_Z$ and $H_{Hyper}$ and (e) in the presence of $H_Z$, $H_{Hyper}$, and $H_Q$. For a given manifold, transitions allowed by the dipolar flip-flop interactions are denoted by red dotted arrows, and their transition rates are indicated as $\Omega_{FF}$. The energy splitting between the states in (d) is only induced by a small hyperfine field (denoted as $\Delta_{HF}$), whereas it is given by both the hyperfine and quadrupole interactions (denoted as $\Delta_{HF+Q}$).

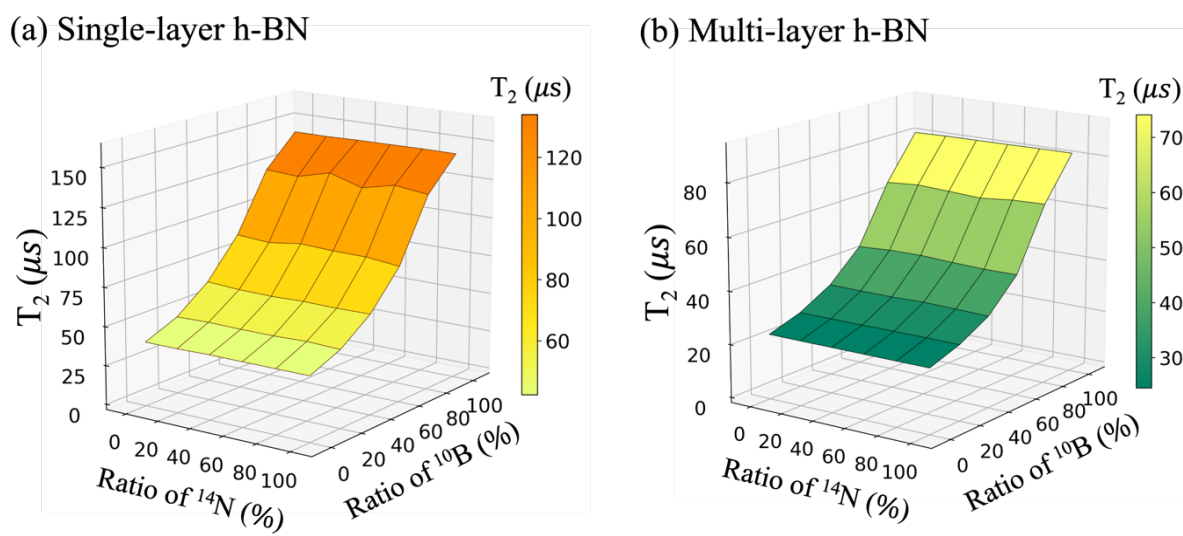

**Figure 2. Enhancement of T$_2$ in h-BN via isotopic engineering (a, b)** Computed T$_2$ of the V$_B^-$ spin in (a) monolayer h-BN and (b) multi-layer h-BN as a function of the ratio of $^{14}$N and $^{10}$B nuclei in the lattice. An external magnetic field of 3 T is applied. The maximum T$_2$ time is obtained when h-BN is 100 % enriched with $^{10}$B both in the single- and multi-layer configurations, yielding a T$_2$ of 143 and 81 μs, respectively.

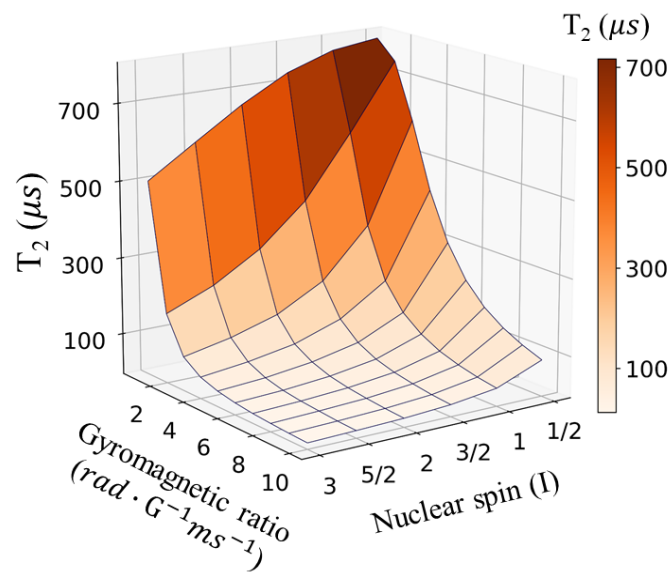

**Figure 3. Hypothetical T$_2$ of V$_B^-$ in h-BN.** Computed T$_2$ as a function of the gyromagnetic ratio and nuclear spin number of $^{10}$B in h-$^{10}$B$^{14}$N. An external magnetic field of 3 T is applied.

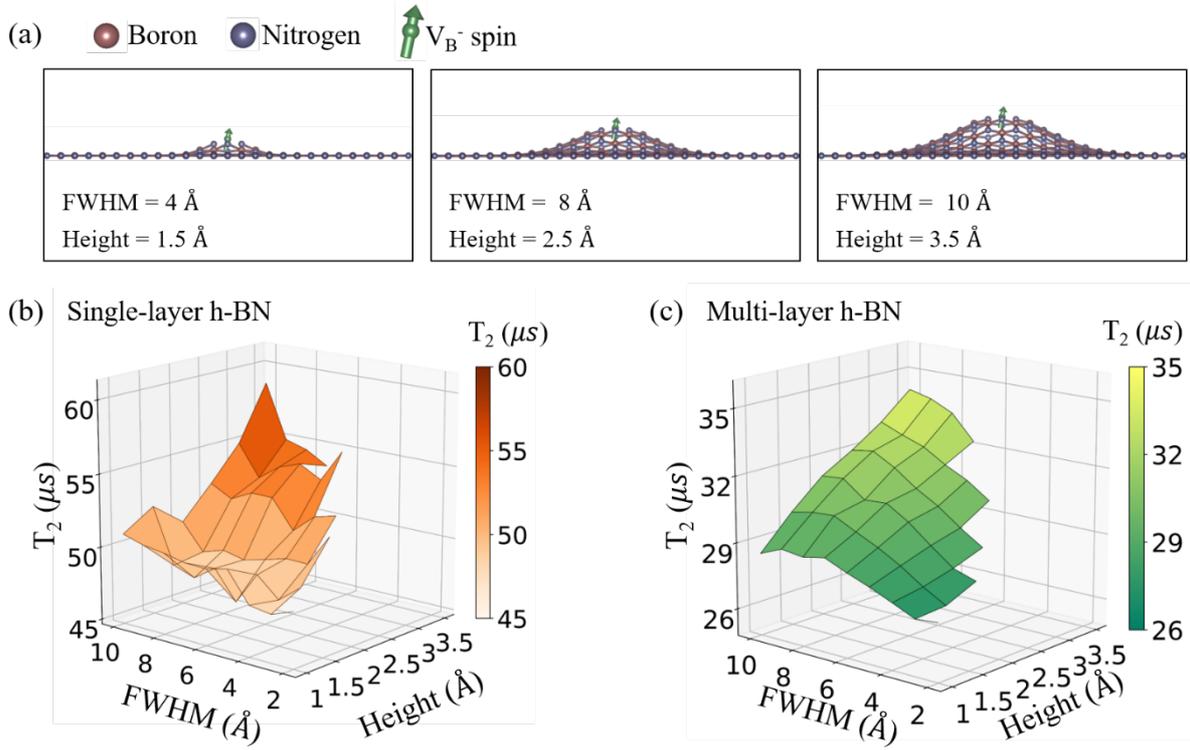

**Figure. 4. Strain-induced increase in $V_B^-$ $T_2$. (a)** Gaussian-shaped lattice deformation of h-BN to induce an inhomogeneous strain around the $V_B^-$. To control the degree of the lattice deformation, the height (Z) and FWHM of the Gaussian function are varied. **(b, c)** Computed $T_2$ times of the $V_B^-$ spin in (b) single-layer and (c) multi-layer h-BN as functions of the FWHM and height of the Gaussian lattice-deformation function. An external magnetic field of 3 T is applied.

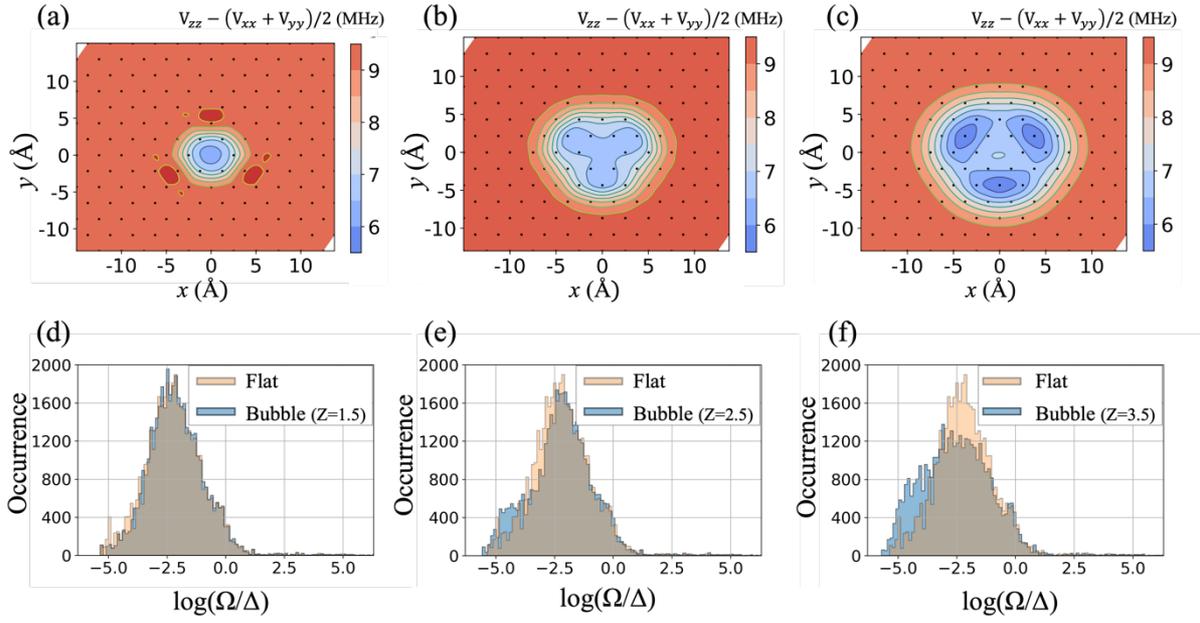

**Figure. 5. Inhomogeneous nuclear quadrupole interaction of $^{10}$B, induced by an inhomogeneous strain. (a–c)** Spatial map of the coefficient of first-order energy shift of the nuclear spins in curved h-BN, due to the quadrupole interaction, which is defined as $\zeta = V_{zz} - (V_{xx}+V_{yy})/2$. Gaussian deformation is applied around the $V_B^-$, and the structures used for (a), (b), and (c) are the same as those shown in Fig 4. (a), characterized by (a) Z = 1.5 Å and FWHW = 4 Å, (b) Z = 2.5 Å and FWHW = 8 Å, and (c) Z = 3.5 Å and FWHW = 10 Å. **(e-f)** Histogram of $\log(\Omega/\Delta)$ of all the possible pseudospins in h-BN with Gaussian lattice deformation, where $\Omega$ is the flip-flop transition rate, and $\Delta$ is the energy gap of a pseudospin (see Supplementary Note 4). The h-BN structures used are characterized by (d) Z = 3.5 Å and FWHW = 10 Å, (e) Z = 1.5 Å and FWHW = 4 Å, and (f) Z = 2.5 Å and FWHW = 8 Å, which are schematically shown in Fig. 4. (a).

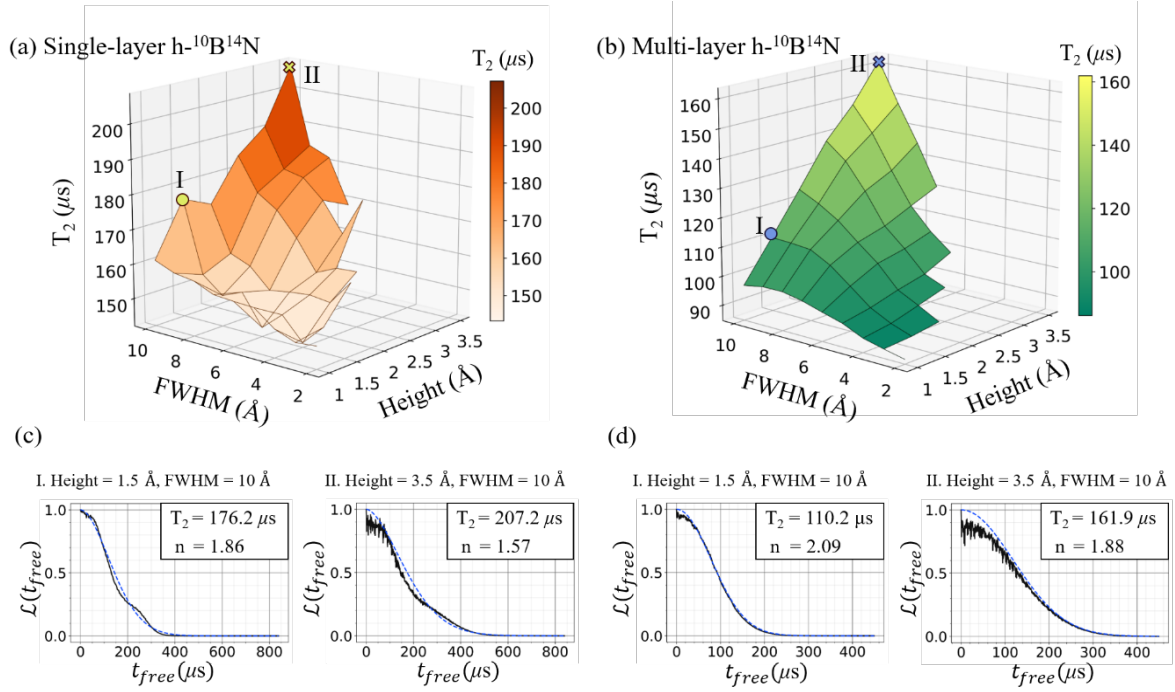

**Figure 6. Maximizing the spin coherence time of $V_B^-$ by combining isotopic and strain engineering. (a, b)** Computed $T_2$ of $V_B^-$ in (a) single-layer and (b) multi-layer h-$^{10}$B$^{14}$N, with Gaussian lattice deformation, as a function of FWHM and height of the Gaussian deformation function. Nitrogen in the lattice is 100 % $^{14}$N. An external magnetic field of 3 T is applied. **(c, d)** $V_B^-$ spin coherence in (c) single-layer and (d) multi-layer h-$^{10}$BN, which has a Gaussian lattice deformation corresponding to condition I (FWHM = 10 Å and height = 1.5 Å) and II (FWHM = 10 Å and height = 3.5 Å) indicated in (a) and (b), respectively.

# TABLE

**Table 1.** Electric field gradients in multi-layered bulk h-BN, and the corresponding quadrupole coupling constants (eQV$_{zz}$) computed using DFT.

| Isotopes | $V_{xx} = V_{yy}$ (eV/Å$^2$) | $V_{zz}$ (eV/Å$^2$) | eQV$_{zz}$ (MHz) | eQV$_{zz}$ (Exp. [57]) (MHz) |
|---|---|---|---|---|
| $^{10}$B | -15.450 | 30.901 | 6.260 | 6.1 |
| $^{11}$B |  |  | 3.004 | 2.9 |
| $^{14}$N | -2.478 | 4.956 | 0.23 | 0.2 |

# MATERIALS & CORRESPONDENCE

*E-mail: hseo2017@ajou.ac.kr

# DATA AVAILABILITY

The data that support the findings of this study are available upon reasonable request to the corresponding author.

# CODE AVAILABILITY

The code that were used in this study are available upon reasonable request to the corresponding author.

ACKNOWLEDGMENTS

This work was supported by the National Research Foundation of Korea (NRF) grant funded by the Korea Government (MSIT) (No. 2018R1C1B6008980, No. 2018R1A4A1024157, No. 2019M3E4A1078666, and No. 2021R1A4A1032085). This work was supported by the Ajou University research fund.

AUTHOR CONTRIBUTIONS

J.L. performed the theoretical calculations. H.P. and J.L. developed the CCE code. H.S. devised and supervised the project. All authors contributed to the data analysis and production of the manuscript.

ADDITIONAL INFORMATION

**Supplementary Information** accompanies the paper.

**Competing interests:** The authors declare no competing interests.